\documentclass[12pt]{iopart}

\usepackage{graphicx}
\usepackage{cite}

\expandafter\let\csname equation*\endcsname\relax
\expandafter\let\csname endequation*\endcsname\relax\usepackage{amsmath}

\usepackage{amsfonts}
\usepackage{color}
\usepackage{textcomp}

\newcommand{\zhat}{\hat{z}}
\newcommand{\etahat}{\hat{\eta}}

\newcommand{\ahat}{\hat{a}}
\newcommand{\xihat}{\hat{\xi}}
\newcommand{\dhat}{\hat{d}}
\newcommand{\IFr}{\mathrm{if}}
\newcommand{\du}{d}
\newcommand{\im}{i}
\newcommand{\ex}{e}
\newcommand{\therm}{\mathrm{th}}
\newcommand{\rr}{\mathrm{rr}}
\newcommand{\bb}{\mathrm{bb}}

\newcommand{\ext}{\mathrm{ext}}
\newcommand{\n}{\mathrm{n}}
\newcommand{\m}{\mathrm{m}}
\newcommand{\p}{\mathrm{p}}
\newcommand{\s}{\mathrm{s}}
\newcommand{\cc}{\mathrm{c}}
\newcommand{\xx}{xx}
\newcommand{\xy}{xy}
\newcommand{\yy}{yy}
\newcommand{\inrm}{\mathrm{in}}

\newcommand{\SiN}{Si$_3$N$_4$}
\newcommand{\wiff}{$\omega_{\text{if}}/2\pi$}

\begin{document}

\title{Cryogenic Optomechanics with a \SiN~Membrane and Classical Laser Noise}

\author{A. M. Jayich$^{1\dagger}$, J. C. Sankey$^{2\dagger}$,  K. B{\o}rkje$^{1\dagger}$, D. Lee$^1$, C. Yang$^1$, M. Underwood$^1$, L. Childress$^1$, A. Petrenko$^1$, S. M. Girvin$^{1,3}$, J. G. E. Harris$^{1,3}$}
\address{$^1$Yale University Department of Physics, 217 Prospect St, New Haven, CT 06511, USA}
\address{$^2$McGill University Department of Physics, 3600 rue University, Montreal, QC H3A2T8, Canada}
\address{$^3$Yale University Department of Applied Physics, 15 Prospect St, New Haven, CT 06511 USA}
\address{$^\dagger$These authors contributed equally to this work.}
\ead{\mailto{jack.sankey@gmail.com}}
\date{\today}

\begin{abstract}
We demonstrate a cryogenic optomechanical system comprising a flexible \SiN~membrane placed at the center of a free-space optical cavity in a 400 mK cryogenic environment. We observe a mechanical quality factor $Q > 4\times 10^6$ for the 261-kHz fundamental drum-head mode of the membrane, and a cavity resonance halfwidth of 60 kHz. The optomechanical system therefore operates in the resolved sideband limit. We monitor the membrane's thermal motion using a heterodyne optical circuit capable of simultaneously measuring both of the mechanical sidebands, and find that the observed optical spring and damping quantitatively agree with theory. The mechanical sidebands exhibit a Fano lineshape, and to explain this we develop a theory describing heterodyne measurements in the presence of correlated classical laser noise. Finally, we discuss the use of a passive filter cavity to remove classical laser noise, and consider the future requirements for laser cooling this relatively large and low-frequency mechanical element to very near its quantum mechanical ground state.
 
\end{abstract}

\maketitle

\section{Introduction}

Cavity optomechanical systems offer a new arena for studying nonlinear optics, the quantum behavior of massive objects, and possible connections between quantum optics and condensed matter systems \cite{2007_OpticsExpress_Kippenberg_Vahala, Marquardt2009Optomechanics, Heinrich2011Collective, Pikovski2012Probing, Isart2011Large, Chang2009Cavity}. Many of the scientific goals for this field share two prerequisites: cooling a mechanical mode close to its ground state, and detecting its zero-point motion with an adequate signal-to-noise ratio. 

The first experiment to satisfy these prerequisites used a conventional dilution refrigerator to cool a piezoelectric mechanical element coupled to a superconducting qubit \cite{OConnell2010Quantum}. The base temperature of the refrigerator ensured that one of the higher-order vibrational modes (a dilatational mode with resonance frequency $\sim$ 6 GHz) was in its quantum mechanical ground state. At the same time, the mechanical element was strongly coupled to a superconducting qubit via its piezoelectric charge, ensuring that the presence of a single phonon in the dilatational mode could be detected with high fidelity. 

Despite the success of this approach, many optomechanics experiments would benefit from the use of low-order mechanical modes, mechanical modes with higher quality factors $Q$ (the mechanical element used in Ref. \cite{OConnell2010Quantum} had $Q \sim 260$), and direct coupling between the mechanical element and the electromagnetic field (i.e., rather than via a qubit). In addition, some experiments will require the mechanical system to couple to optical frequencies (i.e., visible and near-infrared light) \cite{Stannigel2010Optomechanical} in addition to microwaves \cite{Regal2011From}.

A number of groups have developed optomechanical systems in which a high-quality, low-order vibrational mode of an object is coupled to a microwave or optical cavity of very low loss \cite{2007_OpticsExpress_Kippenberg_Vahala, Marquardt2009Optomechanics}. These high-quality-factor mechanical devices typically resonate at frequencies far too low to be cooled to the ground state by conventional refrigeration techniques. Nevertheless, their vibrational modes can be cooled well below the ambient temperature using coherent states of the electromagnetic field (produced, e.g., by an ideal, noiseless laser) \cite{Rae2007Theory, Marquardt2007Quantum}. The technique of using coherent laser light to reduce the temperature of another system (i.e. ``laser cooling'') has been used with great success in the atomic physics community to both prepare a single trapped ion in its motional ground state \cite{Diedrich1989Laser} and provide one of the cooling stages necessary to achieve Bose Einstein condensation in a dilute atomic gase\cite{Wieman1999Atom}. Laser cooling also has a long history in optomechanics, and a number of descriptions of laser-cooled optomechanical systems have been presented in the literature \cite{Braginsky1970Invesitgation, Braginsky1967Ponderomotive, 2007_OpticsExpress_Kippenberg_Vahala, Marquardt2009Optomechanics}.

To date, two groups have described experiments in which laser cooling (or its microwave analog) has been used to reduce the vibrations of a solid object close to its quantum mechanical ground state (i.e., to mean phonon number less than unity) \cite{Teufel2011Sideband,Chan2011Laser}. In these experiments the electromagnetic drive provided both the cooling and single-sideband readout of the mechanical motion.

To achieve a mean phonon number very close to zero, a number of technical obstacles must be overcome. In general, laser cooling is optimized when the mechanical mode is weakly coupled to its thermal bath and well coupled to an electromagnetic cavity. This can be achieved by using a mechanical oscillator of high $Q$, and by applying a strong drive to an optical cavity of high finesse $F$. However even when these criteria are met, there is a minimum temperature that can be achieved by laser cooling. For a laser without any classical noise, this limit is set by the quantum fluctuations of the light in the cavity. Also, as described in Refs. \cite{Rae2007Theory} and \cite{Marquardt2007Quantum}, a laser without classical noise can achieve ground state cooling only if the optomechanical system is in the resolved sideband regime (i.e., the mechanical frequency is larger than the cavity loss rate). However if the laser that is driving the cavity exhibits classical fluctuations, its cooling performance will be degraded because classical fluctuations carry a non-zero entropy \cite{Rabl2009Phasenoise, Diosi2008Laser}. Qualitatively speaking, the fluctuating phase and amplitude of the light result in fluctuating radiation pressure inside the cavity, which in turn leads to random motion of the mechanical element that is indistinguishable from thermal motion. This point has been discussed in the optomechanics literature, and may play an important role in some experiments \cite{Kippenberg2011Phase}.

Here we present a description of an experiment that meets many of the criteria for ground state laser cooling and detection (in that a high quality mechanical element is coupled to a high-finesse cavity in a cryogenic environment), but whose cooling performance is limited by classical laser noise. This experiment employs a membrane-in-the-middle geometry \cite{2008_Nature_Thompson_Harris}, in which a flexible dielectric membrane is placed inside a free-space optical cavity. The typical dimensions of free-space optical cavities lead to the requirement that the membrane have a lateral dimension $\sim$ 1 mm to avoid clipping losses at the beam waist. This leads to a fundamental drum-head mode with a resonance frequency $\sim 10^5$ Hz, requiring laser cooling to $\sim$ 1 \textmu K in order to reach the ground state. Despite this low temperature, this type of optomechanical system is appealing for a number of reasons. The \SiN~ membranes used here exhibit exceptionally high quality factors $Q$ (even when they are patterned into more complex shapes \cite{KimblePersonal}), low optical absorption \cite{Sankey2010Strong}, and compatibility with monolithic, fiber-based optical cavities \cite{FlowersJacobs2012FiberCavityBased}. Furthermore, the membrane-in-the-middle geometry provides access to different types of optomechanical coupling that may serve as useful tools for addressing quantum vibrations \cite{2008_Nature_Thompson_Harris, Sankey2010Strong, 2008_NJP_Jayich_Harris}. 

At a cryogenic base temperature of 400 mK, we observe a mechanical quality factor $Q > 4\times 10^6$ for the 261-kHz fundamental membrane mode, and a cavity resonance halfwidth of 60 kHz, meaning the system operates in the resolved sideband limit. We monitor the membrane's thermal motion using a heterodyne optical circuit capable of simultaneously measuring both of the mechanical sidebands, and find that the observed optical spring and damping quantitatively agree with theory.

To quantify the role of classical laser noise in this system, as well as optomechanical systems more generally, we also present a detailed theoretical model of optomechanical systems that are subject to classical laser noise. This model describes the roles of amplitude noise, phase noise, and amplitude-phase correlations in the multiple beams that are typically used to cool and measure an optomechanical system. Expressions are derived for the heterodyne spectrum expected for optomechanical systems in the presence of correlated noise sources, and we discuss the limits that classical laser noise imposes on cooling and reliably measuring the mean phonon number.

\section{Cryogenic Apparatus}

\begin{figure}
  \includegraphics[width=\textwidth]{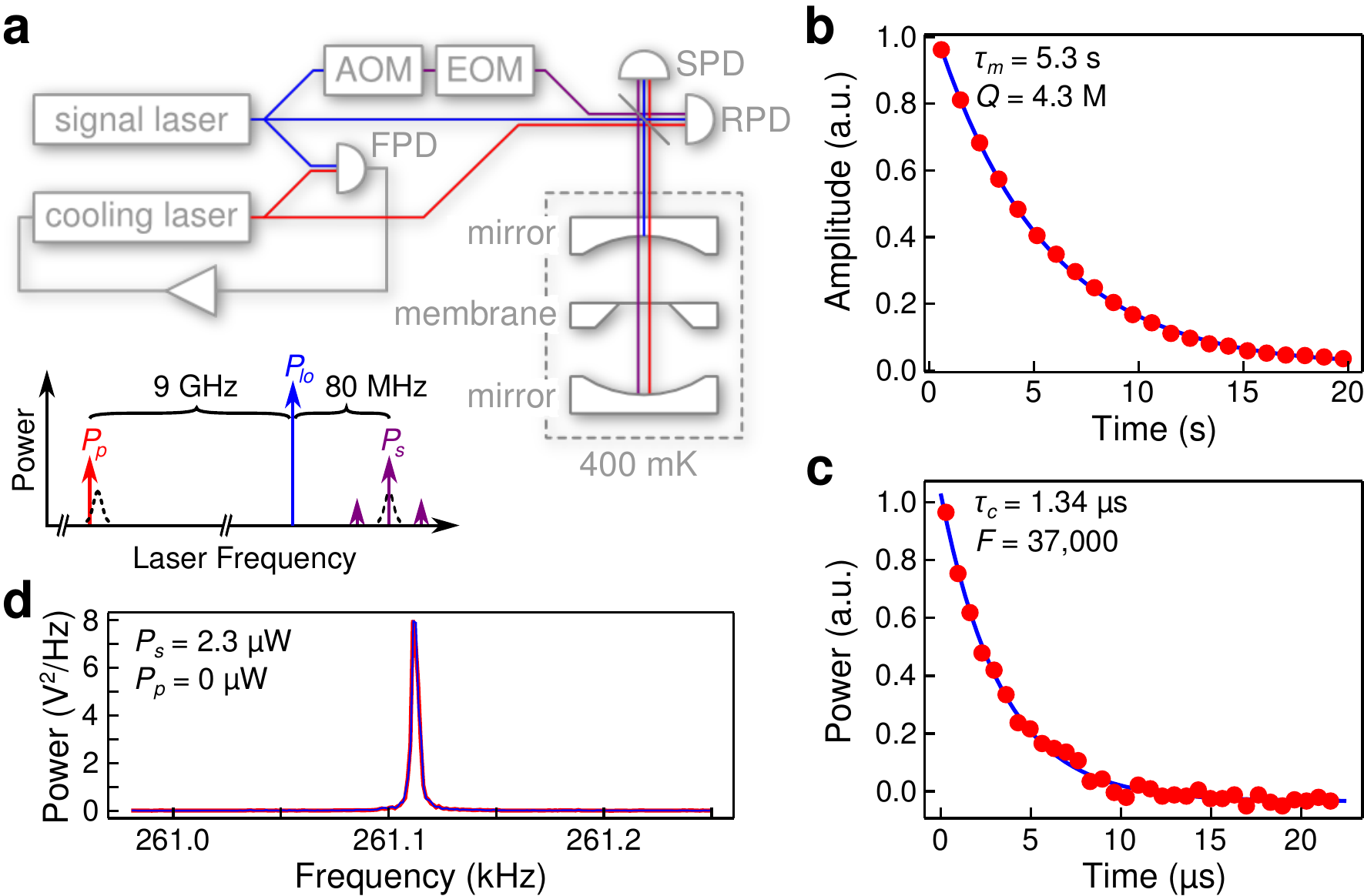}
  \caption{Cryogenic optomechanical system. \
  \textbf{a} Two Nd-YAG lasers probe and cool the cryogenic optomechanical system. The lasers are frequency-locked $\sim$ 9 GHz apart by feeding back on the beat signal from a fast photodiode (FPD). The majority of the signal laser's output serves as a heterodyne local oscillator. The rest is shifted 80 MHz with an acousto-optical modulator (AOM) and then phase modulated using an electro-optical modulator (EOM) with 22\% of the power in $\pm$15 MHz sidebands. These beams land on a sampler, and a small amount is sent to the cold cavity. The remainder lands on a ``reference'' photo diode (RPD) to monitor the heterodyne phase. Light leaving the cavity is collected by another ``signal'' photodiode (SPD) to monitor the membrane's motion. The signal laser is locked to the cavity with the Pound-Drever-Hall (PDH) method using the 15 MHz sidebands. The frequency and amplitude of the cooling laser are fine-tuned with an additional AOM (not shown). 
  \textbf{b} Mechanical ringdown measurement, showing the membrane's amplitude after a drive piezo is turned off. 
  \textbf{c} Cavity ringdown measurement, showing power leaving the cavity after the drive laser is turned off.  
  The solid lines in \textbf{b} and \textbf{c} show exponential fits to the data.
  \textbf{d} Power spectral density of the heterodyne sidebands from the membrane's Brownian motion at 400 mK. The frequency is plotted relative to \wiff = 80 MHz, and the lower sideband (red) has been folded on top of the upper sideband (blue) for comparison. \
}
  \label{fig1}
\end{figure}

Figure \ref{fig1}a shows a schematic of our cryogenic optomechanical system. A 1.5 mm $\times$ 1.5 mm $\times$ 50 nm stoichiometric \SiN~membrane resides at the center of a (nominally) $3.39$ cm long optical cavity. The membrane is mounted on a three-axis cryogenic actuator allowing us to tilt the membrane about two axes and displace it along the cavity axis. The cavity, membrane, and a small set of guiding optics are cooled to approximately 400 mK in a $^3$He cryostat. Free-space laser light is coupled to the cavity via one of the cryostat's clear-shot tubes. 

The most reliable way to measure the membrane's mechanical quality at 400 mK is to perform a mechanical ringdown by driving the membrane at its resonant frequency ($\omega_\text{m}= 2\pi \times 261.15$ kHz) to large amplitude with a nearby piezo, shutting off the drive, and monitoring the decay of the membrane's vibrations. We monitor the membrane's motion interferometrically using a laser of wavelength 935 nm, which is far enough from the design wavelength of our cavity mirror coatings (1064 nm) that the cavity finesse is $\sim 1$; this ensures the measurement exerts no significant back action upon the membrane. Figure \ref{fig1}b shows a typical mechanical ringdown measurement. To ensure the membrane motion is in the linear regime, we let it ring down until its frequency stabilizes before fitting the data to an exponential curve (the inferred time constant is then insensitive to the choice of time window). The observed ringdown time $\tau_\text{m} = 5.3$ s corresponds to a mechanical quality factor $Q$ = 4.3 million at 400 mK, though this value varies with thermal cycling (i.e. between 400 mK and 4 K), and typically ranges from $\sim 4-5$ million.

As shown in Fig. \!\!\!\!\!\ref{fig1}a, two independent Nd-YAG lasers (wavelength $\lambda$ = 1064 nm) provide a total of five beams for driving the cavity and performing the heterodyne detection of the membrane's motion (described below). To achieve a large optomechanical back action with these lasers, we require a high-finesse optical cavity. The top and bottom mirrors in Fig. \!\ref{fig1}a are designed to have a power reflectivity exceeding 99.98\% and 99.998\% respectively at $\lambda = 1064$ nm, which would correspond to a cavity finesse of 30,000. Generally these mirrors perform above this specification, however. Figure \ref{fig1}c shows the results of a typical cavity ringdown measurement performed by toggling the power of a laser driving the cavity, and collecting the power leaking out of the cavity when the drive is shut off. The measured time constant $\tau_c = $ 1.34 \textmu s corresponds to a finesse of $F=37,000$. This value generally depends on the day the data was taken and the orientation of the membrane. It is lower than the value we measured after initially cooling to 400 mK ($\sim 80,000$). We believe this reduction was caused by either gradual condensation of materials on the surfaces over months of operation, or a change in the membrane's alignment, which can steer the cavity mode away from a high-performance region of the end mirrors (a spatial dependence of cavity-mirror performance was also observed in Ref. \cite{Sankey2010Strong}). The finesse measured in Fig. \ref{fig1}c corresponds to a cavity loss rate of $\kappa/2\pi$~= 120 kHz, meaning the cryogenic optomechanical system operates in the resolved sideband regime, a condition necessary for ground-state cooling \cite{Rae2007Theory,Marquardt2007Quantum}.

The first purpose of this apparatus is to perform a heterodyne measurement of the membrane's motion. As shown in Fig. \ref{fig1}a, light from the ``signal laser'' is split into several frequencies before it interacts with the cavity. The inset of Fig. \ref{fig1}a shows a summary of the relative magnitudes and frequencies of the laser light landing on the cavity, with dashed lines roughly illustrating the susceptibility of the different cavity resonances. Most of the light serves as a local oscillator tuned far from the cavity resonance; this power $P_{\text{lo}}$ simply bounces off the first cavity mirror and returns to a ``signal'' photodiode (SPD). A small fraction of this light is shifted by \wiff~= 80 MHz using an acousto-optical modulator (AOM) and is used to both lock the laser near the cavity resonance and record the membrane's motion. Locking is achieved via the Pound-Drever-Hall technique \cite{Black2001Introduction} with 15 MHz sidebands generated by an electro-optical modulator (EOM). A sampler directs $\sim 5$\% of these beams' power into the cryostat and cavity. We use the remaining 95\% (sent to a ``reference'' photodiode RPD) to monitor the laser's phase and power. The sampler then passes $\sim 95$\% of the light escaping the cryostat through to the signal photodiode. This signal is demodulated at the beat note \wiff~= 80 MHz in order to simultaneously detect the two sidebands generated by the membrane's thermal motion. 

Figure \ref{fig1}d shows a typical power spectral density of these sidebands. A peak appears at the membrane's fundamental mechanical frequency $\omega_\text{m}/2\pi \approx 261.1$ kHz as expected. The sidebands are identical, as expected for an interferometric measurement in which the laser noise contributes a negligible amount of force noise compared to the thermal bath and the mean phonon number is $\gg1$.

The second purpose of this apparatus is to manipulate the membrane with optical forces, and so we include a second (cooling / pump) laser that addresses a different longitudinal mode of the cavity. If the cooling and signal beams address the same cavity mode, the beating between the two beams leads to a large heterodyne signal that clouds our measurement and a strong mechanical drive at the beat frequency (which usually is close to the mechanical frequency). This can cause the system to be unstable and makes the data difficult to interpret. To overcome this challenge, we lock the cooling and signal lasers such that they address different longitudinal cavity modes roughly 9 GHz apart. The longitudinal modes are chosen to be two free spectral ranges apart so that the dependence of cavity resonance frequency on membrane displacement is approximately the same for the two modes. This way, drift or vibrations in the membrane mount will (to lowest order) not change the relative frequencies of the modes. With the lasers locked in this way, any beating between the cooling and signal lasers occurs at frequencies that are irrelevant to the membrane's mechanics. 

As shown in Fig. \ref{fig1}a, the two lasers are locked by picking off a small portion of both beams and generating an error signal based on the frequency of their beat note. We have locked the free-running lasers $\sim 9$ GHz apart with an RMS deviation of $\sim 10$ Hz. When the signal laser is simultaneously locked to the membrane cavity, however, this performance degrades to an RMS deviation of $\sim 1$ kHz; this is because the membrane cavity is quite sensitive to environmental noise such as acoustic vibrations in the room, which injects additional noise into the signal laser's frequency (this first-generation cryogenic apparatus did not include significant vibration isolation). When the two lasers are locked to each other and the signal laser is locked to the membrane cavity, the cooling laser can then be fine-tuned relative to its cavity mode using an additional AOM (not shown).

\begin{figure}
  \includegraphics[width=\textwidth]{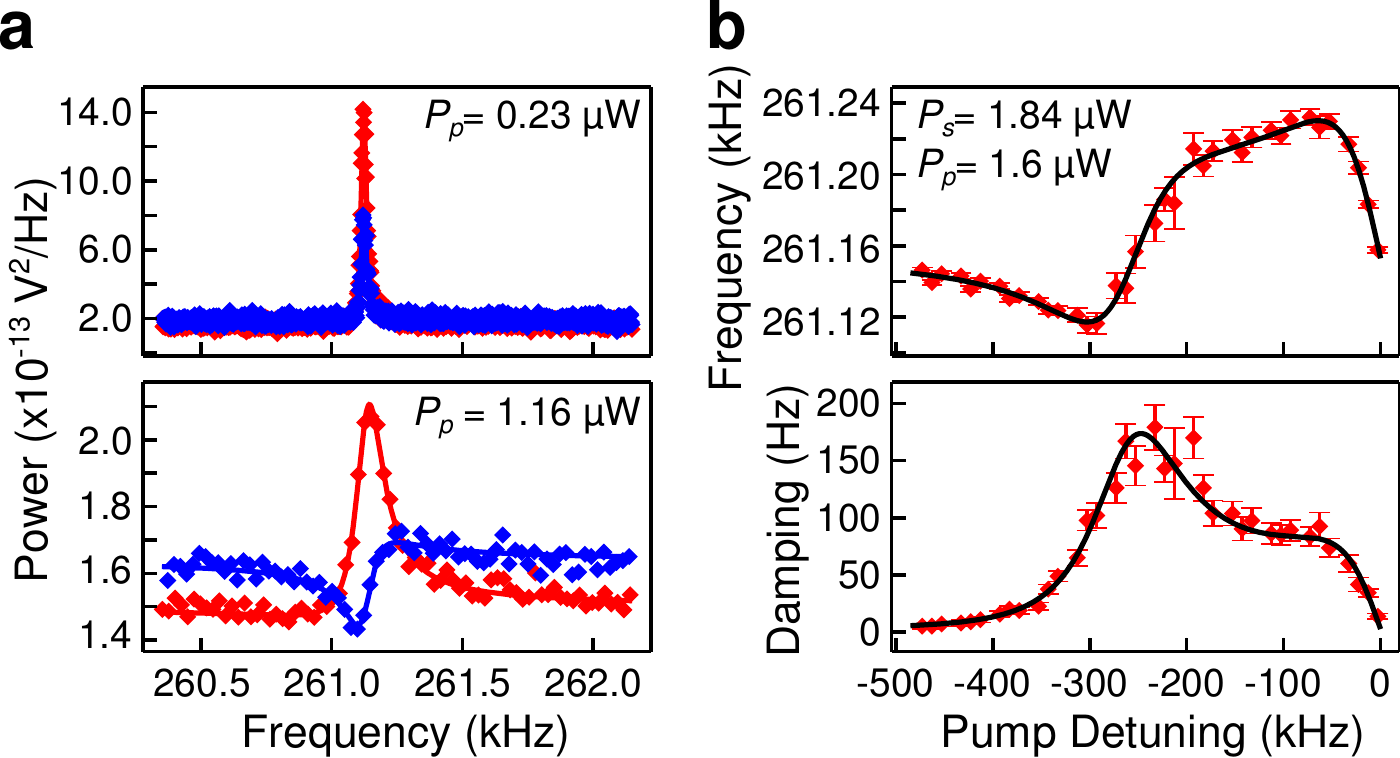}
  \caption{Response of membrane to the cooling laser. \textbf{a} Typical heterodyne spectra (red and blue sidebands folded on top of each other) for increasing values of cooling laser power $P_\text{p}$ at $\Delta_\text{p}/2\pi = -250$ kHz. Solid lines are fits to Fano lineshapes (simultaneously fitting the width and frequency of both sidebands). \textbf{b} Membrane frequency and damping determined from Fano fits (similar to \textbf{a}) for different values of $\Delta_\text{p}/2\pi$. Solid lines represent a simultaneous fit of these two data sets to optomechanical theory.}
  \label{fig2}
\end{figure}

The cooling beam adds a significant optomechanical damping and spring to the membrane, so the linewidth and center frequency of the sidebands in Fig. \ref{fig1}d depend on its detuning $\Delta_\text{p}$ and power $P_\text{p}$. Figure \ref{fig2}a shows typical heterodyne spectra for the cooling beam red-detuned by $\Delta_\text{p}/2\pi = -250$ kHz. As the cooling power $P_\text{p}$ is increased (from $P_\text{p}=0$ in Fig. \ref{fig1}a), the membrane's vibrations are laser cooled; the linewidth increases and the integrated area under the curve decreases qualitatively as expected. At high $P_\text{p}$ the red and blue sidebands exhibit a large asymmetry. We find the spectra are always well-fit by a Fano lineshape. 

Figure \ref{fig2}b shows the membrane's mechanical frequency and damping as a function of $\Delta_\text{p}$. We simultaneously fit the frequency and damping to the theory described in Ref. \cite{Marquardt2007Quantum} (and outlined in section \ref{section-model} below), allowing four parameters to vary: the free spectral range FSR, the ratio between the cavity's loss through the entrance mirror to the total cavity loss $\kappa_{\text{ext}}/\kappa$, the bare mechanical frequency $\omega_\text{m}$, and and the signal beam detuning $\Delta_\text{s}$. The results of this fit are: FSR $= 8.7673410 \text{ GHz} \pm 5 \text{ kHz}$ (Note the statistical fit error was 460 Hz. The quoted error reflects the precision of a frequency measurement used to generate 9 GHz error signal.), $\kappa_{\text{ext}}/\kappa = 0.243 \pm 0.003$, $\omega_\text{m}/2\pi = 261150.3 \pm 0.9 $ Hz, and $\Delta_\text{s}/2\pi = -880 \pm 250$ Hz. The precise value of the FSR adds an overall offset to $\Delta_\text{p}$ (i.e. a horizontal shift in Fig. \ref{fig2}b). The estimate of FSR from this fit is significantly more precise than our independent estimate of 8.84 GHz based on cavity length. The ratio $\kappa_{\text{ext}}/\kappa$ simultaneously scales the optical spring and damping strength. This can be independently estimated as $\kappa_{\text{ext}}/\kappa = 0.2$ from measuring the cavity ringdown and the fraction of the incident light lost in the cavity with the laser tuned on resonance (64 \% in this case). This estimate is lower than the fit value by 20\%, which we attribute to imperfect cavity mode matching and that the membrane position varies by $\sim 10$ nm  during measurements, which can affect the cavity finesse \cite{Sankey2010Strong}. We allow the bare mechanical frequency to float because we find that it can drift by a few Hz on the hour time scale. This adds a constant offset to the frequency plot in Fig. \ref{fig2}b. Finally, for this particular experiment we locked the signal beam as close to resonance as possible, but as this tends to drift on the scale of hours, we left $\Delta_\text{s}$ as a fitting parameter. $\Delta_\text{s}$ is responsible for adding a very small constant offset to the damping and spring. All other parameters such as the cavity finesse and input power were measured independently. The simultaneous fit is thus heavily constrained and agrees with the data very well. We also find that the fit is similarly convincing if we simply fix $\Delta_\text{s} = 0$ and $\omega_\text{m}/2\pi = 261.15$ kHz (a typical value of $\omega_\text{m}$).

While the optical spring and damping in Fig. \ref{fig2}b are well-modeled by standard theory, the interpretation of the sideband amplitudes and lineshapes in Fig. \!\!\ref{fig2}a is not obvious. As we now discuss, the Fano lineshape arises from interference between the membrane's response to classical laser noise and the classical laser noise itself, an effect similar to what is seen in single-sideband measurements in other optomechanical systems \cite{Rocheleau2010Preparation,Teufel2011Sideband,Gavartin2012Hybrid}.

\section{General Model of Optomechanics with Classical Laser Noise}
\label{section-model}

In this section, we present and solve the equations of motion for the optical cavity and the mechanical oscillator. Since the local oscillator beam is far off any cavity resonance frequency, we can neglect it here. We will let $\ahat_\s$ be the bosonic annihilation operator of the cavity mode addressed by the lock/signal beam, whereas $\ahat_\p$ is the annihilation operator for the cavity mode addressed by the cooling (pump) beam. The position operator of the mechanical oscillator is $\hat{x} = x_0 + x_\mathrm{zpf} \left(\hat{c} + \hat{c}^\dagger\right)$, where $\hat{c}$ is the phonon annihilation operator, $x_0 = \langle \hat{x} \rangle$ and $x_\mathrm{zpf}$ is the size of the zero point fluctuations. The Hamiltonian is
\begin{equation}
  \label{eq:Hamiltonian}
  H = \sum_{j = \s,\p} \hbar \left(\omega_j + g_j \hat{x} \right) \ahat^\dagger_j \ahat_j + \hbar \omega_\m \hat{c}^\dagger \hat{c} + H_\mathrm{drive} + H_\mathrm{diss}
\end{equation}
The interaction term describes the modulation of the cavity resonance frequencies by the motion of the mechanical oscillator, $H_\mathrm{drive}$ describes the laser drive and $H_\mathrm{diss}$ describes the coupling to both the electromagnetic and mechanical environment. This coupling to external degrees of freedom is conveniently described by input-output theory \cite{Collett1984Squeezing,Clerk2010Introduction}, which gives rise to the equations of motion
\begin{eqnarray}
  \label{eq:EOM}
  \dot{\ahat}_j  & = & -\left(\frac{\kappa_j}{2} + \im \omega_j \right) \ahat_j - \im g_j \hat{x} \ahat_j + \sqrt{\kappa_{j,\ext}} \, \ahat_{j,\inrm} + \sqrt{\kappa_{j,\mathrm{int}}} \, \xihat_{j} \quad , \quad j = \s,\p \\
  \dot{\hat{c}} & = & - \left(\frac{\gamma}{2} + \im \omega_\m \right) \hat{c} - \im \sum_j g_j \ahat^\dagger_j \ahat_j + \sqrt{\gamma} \, \etahat \ .
\end{eqnarray}
Here, $\kappa_{j,\mathrm{ext}}$ is the decay rate of mode $j$ through the mirror which couples the cavity to the external laser drive, whereas $\kappa_{j,\mathrm{int}}$ describes other types of optical decay. The total linewidth of cavity mode $j$ is $\kappa_j = \kappa_{j,\mathrm{ext}} + \kappa_{j,\mathrm{int}}$. The input modes $\xihat_{j}$ describe optical vacuum noise and fulfill $\langle \xihat_{j}(t) \xihat^\dagger_{j'}(t') \rangle = \delta(t-t') \delta_{j,j'}$ and $\langle \xihat^\dagger_{j}(t) \xihat_{j'}(t') \rangle = 0$. The coupling to the laser drive is described by the input mode
\begin{equation}
  \label{eq:Input}
  \ahat_{j,\inrm}(t) = \ex^{-\im \Omega_{j} t} \left[K_j + \frac{1}{2} \left(\delta x_j(t) + \im \, \delta y_j(t) \right) \right] + \xihat_{j,\inrm} 
\end{equation}
where $K_j = \sqrt{P_j/\hbar \Omega_j}$, with $\Omega_\s$ ($\Omega_\p$) being the drive frequency and $P_\s$ ($P_\p$) the power of the lock (cooling) beam. We have introduced the classical variables $\delta x_j$ and $\delta y_j$ which describe technical laser amplitude and phase noise, respectively. Since we will only be concerned with the noise close to the mechanical frequency $\omega_\m$, we can assume a white noise model where 
\begin{eqnarray}
  \label{eq:ClassNoise}
  \langle \delta x_j(t) \delta x_{j'}(t') \rangle & = & C_{j,\xx} \delta(t-t') \delta_{j,j'} \\
  \langle \delta y_j(t) \delta y_{j'}(t') \rangle & = & C_{j,\yy} \delta(t-t')  \delta_{j,j'}  \nonumber \\
  \langle \delta x_j(t) \delta y_{j'}(t') \rangle & = & C_{j,\xy} \delta(t-t') \delta_{j,j'} \nonumber  
\end{eqnarray}
The amplitude and phase noise is characterized by the real numbers $C_{j,\xx}, C_{j,\yy} \geq 0$ and $C_{j,\xy}$ that are proportional to laser power. The Cauchy-Bunyakovsky-Schwarz inequality dictates that $C_{j,\xy}^2 \leq C_{j,\xx} C_{j,\yy}$. Note that $C_{j,\xx} = 1$ or $C_{j,\yy} = 1$ corresponds to the condition in which the laser's classical noise is equal to its quantum noise. The operator $\xihat_{j,\inrm}$ describes vacuum noise and obeys the same relations as $\xihat_{j}$. The intrinsic linewidth of the mechanical oscillator is $\gamma$, and $\etahat$ describes thermal noise obeying $\langle \etahat(t) \etahat^\dagger(t') \rangle \approx \langle \etahat^\dagger(t) \etahat(t') \rangle = n_\therm \delta(t-t')$, where $n_\therm \approx k_\mathrm{B}T/\hbar \omega_\m$ is the phonon number in the absence of laser driving.

For sufficiently strong driving and weak optomechanical coupling, we can linearize the equations of motion by considering small fluctuations around an average cavity amplitude. We write
\begin{equation}
  \label{eq:MeanFluct}
  \ahat_j(t) = \ex^{-\im \Omega_{j} t} \left( \bar{a}_j + \dhat_j(t) \right)
\end{equation}
where
\begin{equation}
  \label{eq:Mean}
  \bar{a}_j = \frac{\sqrt{\kappa_{j,\ext}} \, K_j}{\kappa_j/2 - \im \Delta_j}
\end{equation}
and $\Delta_j = \Omega_j - \omega_j - g_j x_0$ is the laser detuning from the cavity resonance in the presence of a static membrane. Defining the dimensionless position operator $\zhat = \hat{c} + \hat{c}^\dagger$, the Fourier transform as $f^{(\dagger)}[\omega] = \int_{-\infty}^{\infty} \du t \, \ex^{\im \omega t} f^{(\dagger)}(t)$, and the susceptibilities
\begin{equation}
  \label{eq:Susceptibilities}
  \chi_{j,\cc}[\omega] = \frac{1}{\kappa_j/2 - \im (\omega + \Delta_j)} \quad , \quad \chi_\m[\omega] = \frac{1}{\gamma/2 - \im (\omega - \omega_\m)} \ ,
\end{equation}
the solution to the linearized equations can be expressed as

\begin{eqnarray}
  \label{eq:SolutionEOM}
  \dhat_j[\omega] & = & \chi_{j,\cc}[\omega] \Big(\zeta_j[\omega] - \im \alpha_j \hat{z}[\omega] \Big) \\ 
  \hat{z}[\omega] & = & \frac{1}{N[\omega]} \Bigg[\sqrt{\gamma} \left( \chi_\m^{-1 \, \ast}[-\omega] \eta[\omega] + \chi_\m^{-1}[\omega] \eta^\dagger[\omega] \right) \notag\\
  &&- 2 \omega_\m \sum_j \left( \alpha_j^\ast \chi_{j,\cc}[\omega] \zeta_j[\omega] + \alpha \chi_{j,\cc}^\ast[-\omega] \zeta^\dagger[\omega] \right) \Bigg] \ . 
\end{eqnarray}

We have introduced the effective coupling rates $\alpha_j = g_j x_\mathrm{zpf} \bar{a}_j$, the operators
\begin{equation}
  \label{eq:zetadef}
  \zeta_j[\omega] = \sqrt{\kappa_{j,\ext}} \left[ \frac{1}{2} \left(\delta x_j[\omega] + \im \delta y_j[\omega] \right) + \xihat_{j,\inrm}[\omega] \right] + \sqrt{\kappa_{j,\mathrm{int}}} \, \xihat_{j}[\omega] \ ,
\end{equation}
and the function 
\begin{equation}
  \label{eq:Ndef}
  N[\omega] = \chi^{-1}_\m[\omega] \chi^{-1 \, \ast}_\m[-\omega] - 2 \im \omega_\m \sum_j |\alpha_j|^2 \left( \chi_{j,\cc}[\omega] - \chi^\ast_{j,\cc}[-\omega] \right) \ .
\end{equation}
Eqs.~\eqref{eq:SolutionEOM} gives the optical output field $\hat{a}_{j,\mathrm{out}}(t) = \sqrt{\kappa_{j,\mathrm{ext}}} \, \ahat_j(t) - \ahat_{j,\inrm}(t)$ from mode $j$.

For later use, we calculate the average phonon number $n_\m = \langle \hat{c}^\dagger \hat{c} \rangle$. In the weak coupling limit $|\alpha_\s|, |\alpha_\p| \ll \kappa_\s, \kappa_\p$, one finds
\begin{equation}
  \label{eq:19}
  n_\m = \frac{\gamma n_\therm + \sum_j \gamma_{j} n_{j}}{\tilde{\gamma}} \ .
\end{equation}
Here, $\tilde{\gamma} = \gamma + \gamma_\s + \gamma_\p$ is the effective mechanical linewidth, and the optical contributions to it are given by
\begin{equation}
  \label{eq:gammaOpt}
  \gamma_{j} =   -4 |\chi_{j,\cc}[\omega_\m]|^2 |\chi_{j,\cc}[-\omega_\m]|^2 \Delta_j |\alpha_j|^2 \kappa_j \, \omega_\m \ .
\end{equation}
Furthermore, we define

\begin{eqnarray}
  \label{eq:23}
  \gamma_{j} n_{j} &=& \frac{|\alpha_j|^2}{4} \Big\{ \kappa_{j,\mathrm{ext}} \Big[|B_{j,+}[\omega_\m]|^2 C_{j,\xx} + |B_{j,-}[\omega_\m]|^2 C_{j,\yy} \notag \\
  & & + 2 \, \mathrm{Im} (B_{j,+}[\omega_\m] B^\ast_{j,-}[\omega_\m]) C_{j,\xy} \Big] + \kappa_j |\chi_{j,\cc}[-\omega_\m]|^2 \Big\} \ 
\end{eqnarray}
with $B_{j,\pm}[\omega] = \ex^{-\im \phi_j} \chi_{j,\cc}[\omega] \pm \ex^{\im \phi_j} \chi^\ast_{j,\cc}[-\omega]$ and $\ex^{\im \phi_j} = \alpha_j/|\alpha_j|$. Finally, we also note that the optical spring effect leads to an effective mechanical resonance frequency $\tilde{\omega}_\text{m} = \omega_\text{m} + \delta_\s + \delta_\p$, where
\begin{equation}
  \label{eq:deltaOmegam}
  \delta_j = 2 |\chi_{j,\cc}[\omega_\m]|^2 |\chi_{j,\cc}[-\omega_\m]|^2 \Delta_j |\alpha_j|^2 [(\kappa_j/2)^2 - \omega_\m^2 + \Delta_j^2] 
\end{equation}
is the shift due to mode $j$.
  
\section{Toy example}

To illustrate the role of technical noise in the optical sidebands, we consider a simplified example. We treat the optomechanical system classically, and focus on a single optical mode (omitting the index) with amplitude $a(t) = \ex^{-\im \Omega t} (\bar{a} + d(t))$, where $d(t)$ are the classical fluctuations around a mean amplitude $\bar{a}$. In addition to neglecting vacuum noise, we also neglect laser phase noise and thermal noise of the mechanical bath. Finally, we consider the case where the cavity is driven on resonance, i.e.~$\Delta = 0$. The equations of motion are then
\begin{eqnarray}
  \label{eq:ToyEOM}
  \dot{d} & = & - \frac{\kappa}{2} d - \im \alpha z + \frac{\sqrt{\kappa_\ext}}{2} \delta x(t) \\
  \dot{c} & = & - \left(\frac{\gamma}{2} + \im \omega_\m \right) c - \im \alpha \left(d + d^\ast \right)
\end{eqnarray}
with $\alpha$ real. Instead of considering white amplitude noise, we imagine that the amplitude of the drive is modulated at a frequency $\omega_\n$, such that $\delta x(t) = 2 \sqrt{C_{xx}} \cos \omega_\n t$. The optical force on the oscillator is then proportional to 
\begin{equation}
  \label{eq:ToyForce}
   d(t) + d^\ast(t) = 2 \sqrt{\kappa_\ext C_{xx} }   \, |\chi[\omega_\n]| \cos (\omega_\n t - \vartheta_\n) 
\end{equation}
where the phase $\vartheta_\n$ is defined by $\chi_\cc[\omega_\n] = |\chi_\cc[\omega_\n]| \ex^{\im \vartheta_\n}$. The dimensionless oscillator position becomes
\begin{equation}
  \label{eq:ToyzOpt}
  z(t) = 2 \sqrt{\kappa_\ext C_{xx} } \, \alpha \, |\chi_\cc[\omega_\n]| \Big[ \cos (\omega_\n t - \vartheta_\n) \mathrm{Im} \, \chi_\m[\omega_\n]  - \sin (\omega_\n t - \vartheta_n) \mathrm{Re} \, \chi_\m[\omega_\n] \Big]
\end{equation}
when assuming $\omega_\n$ is positive and close to $\omega_\m$, and $\omega_\m/\gamma \gg 1$. The real part of the mechanical susceptibility is a Lorentzian as a function of $\omega_\n$, whereas the imaginary part is antisymmetric around the mechanical frequency:
\begin{equation}
  \label{eq:ToyRealImag}
  \mathrm{Re} \, \chi_\m[\omega_\n]  = \frac{\gamma/2}{(\gamma/2)^2 + (\omega_\n - \omega_\m)^2} \quad , \quad \mathrm{Im} \, \chi_\m[\omega_\n]  = \frac{\omega_\n - \omega_\m}{(\gamma/2)^2 + (\omega_\n - \omega_\m)^2} \ .
\end{equation} 
As one would expect, the mechanical oscillation goes through a phase shift of $\pi$ as the modulation frequency $\omega_\n$ is swept through the mechanical resonance, and the oscillation is out of phase with the force at resonance $\omega_\n = \omega_\m$. 

We write the optical output amplitude $d_\mathrm{out}(t) = \sqrt{\kappa_\ext} d(t) - \delta x(t)/2$ as a sum of two terms,
\begin{equation}
  \label{eq:ToydOut}
  d_\mathrm{out}(t) = d_{\mathrm{out},\delta x}(t) + d_{\mathrm{out},z}(t) \ ,
\end{equation}
where $d_{\mathrm{out},\delta x}(t)$ is the amplitude for the reflected and cavity filtered signal $\delta x(t)$, whereas $d_{\mathrm{out},z}(t)$ is the part that comes from the motion of the mechanical oscillator. We define the output spectrum as
\begin{equation}
  \label{eq:ToySpectrum}
  S[\omega] = \int_{-\infty}^\infty \du \tau \, \ex^{\im \omega \tau}  \langle d_\mathrm{out}^\ast(t + \tau) d_\mathrm{out}(t) \rangle_\mathrm{time} \ ,
\end{equation}
where $\langle \ \rangle_\mathrm{time}$ denotes averaging over the time $t$.

The spectrum consists of three terms, $S[\omega] = S_{\delta x,\delta x}[\omega] + S_{z,z}[\omega] + S_{\delta x,z}[\omega]$. The first term is the spectrum of $d_{\mathrm{out},\delta x}(t)$, which becomes
\begin{equation}
  \label{eq:ToySxx}
  S_{\delta x,\delta x}[\omega] = \frac{C_{xx}}{4} \big|\kappa_\ext \chi_\cc[\omega_\n]  - 1 \big|^2 \times 2 \pi \Big[\delta(\omega - \omega_\n)  + \delta(\omega + \omega_\n) \Big] \ .
\end{equation}
The absolute value describes the promptly reflected signal, the cavity filtered signal, and their interference. The second term in $S[\omega]$ is the spectrum of $d_{\mathrm{out},z}(t)$, which is proportional to the position spectrum of the mechanical oscillator. We find
\begin{equation}
  \label{eq:ToySzz}
  S_{z,z}[\omega] = \kappa_\ext^2 \alpha^4 C_{xx} |\chi_\cc[\omega_\n]|^4 |\chi_\m[\omega_\n]|^2 \times  2 \pi \Big[\delta(\omega - \omega_\n)  + \delta(\omega + \omega_\n) \Big] \ .
\end{equation}
This is proportional to the absolute square of the mechanical susceptibility, which has a Lorentzian dependence on $\omega_\n$, as one would expect from a damped and driven harmonic oscillator. Note also that $S_{z,z}[\omega]$ is symmetric in $\omega$ as is required of a spectrum of a real, classical variable \cite{Clerk2010Introduction}.

The last term in $S[\omega]$ results from optomechanical correlations between the modulation $\delta x$ and the oscillator position $z$:

\begin{eqnarray}
  \label{eq:ToySxz}
  S_{\delta x,z}[\omega] & \equiv & \int_{-\infty}^\infty \du \tau \, \ex^{\im \omega \tau} \langle d^\ast_{\mathrm{out},z}(t + \tau) d_{\mathrm{out},\delta x}(t)  + d^\ast_{\mathrm{out},\delta x}(t + \tau) d_{\mathrm{out},z}(t) \rangle_\mathrm{time}  \notag\\
  & = & \kappa_\ext \alpha^2 C_{xx} |\chi_\cc[\omega_\n]|^2 \Big[ \left( \kappa_\ext |\chi_\cc[\omega_\n]| \cos \vartheta_\n - \cos 2 \vartheta_\n \right) \mathrm{Re} \, \chi_\m[\omega_\n]  \notag \\
  & &- \left( \kappa_\ext |\chi_\cc[\omega_\n]| \sin \vartheta_\n - \sin 2 \vartheta_\n \right) \mathrm{Im} \, \chi_\m[\omega_\n] \Big] \notag  \\ 
  & & \times 2 \pi \Big[\delta(\omega - \omega_\n)  - \delta(\omega + \omega_\n) \Big] \ .
\end{eqnarray}
We see that this term depends on both the real and imaginary parts of the mechanical susceptibility. Note also that the term $S_{\delta x,z}[\omega]$ is antisymmetric in $\omega$. 

So far we considered amplitude modulation at a single frequency $\omega_\n$. In the case of white noise, there is amplitude modulation at all frequencies simultaneously. The spectrum in that case can be found by simply integrating the above spectrum over all frequencies $\omega_\n$. In the limit where the mechanical decay rate is small compared to the cavity decay rate, $\gamma \ll \kappa$, this gives a spectrum consisting of a noise floor, a Lorentzian $|\chi_\m[\omega]|^2$, and the antisymmetric function given by the imaginary part of the mechanical susceptibility. 

There are two important lessons to be learned from this calculation. The first is that the sidebands of the optical output spectrum are not Lorentzian in general, but can also have an antisymmetric part due to optomechanical correlations. The second is that even if the antisymmetric parts are small or vanish (which for example happens when $\kappa_\ext = \kappa$ in this example) and the two sidebands are Lorentzian, one cannot necessarily conclude that an asymmetry between these peaks at zero detuning is due to the mechanical oscillator being in the quantum regime. An asymmetry between the Lorentzian peaks can also occur due to classical optomechanical correlations. In Section \ref{sec:SidebandWeights}, we will see that neglecting this effect can lead to an underestimation of the effective phonon number.

\section{The heterodyne spectrum}
\label{section-heterodyne}

We now calculate the heterodyne spectrum that results from beating between the local oscillator beam and one of the beams entering the cavity. For this calculation, we need not specify whether it is the lock or the cooling beam that is used for readout. We will simply refer to it as the measurement beam below. To simplify the notation, we will drop the subscript ($\s$ or $\p$) on the operators and parameters that refer to the measurement beam. The other beam will not affect the heterodyne spectrum, except indirectly through the renormalized frequency, linewidth, and mean phonon number of the mechanical oscillator. We can thus omit this beam in the discussion below.

The local oscillator beam is at the frequency $\Omega - \omega_\IFr$, where $\omega_\IFr > 0$ is the intermediate frequency between the measurement beam and the local oscillator. Including the local oscillator, the external input mode is now
\begin{equation}
  \label{eq:InputLO}
  \ahat_{\inrm}(t) = \ex^{-\im \Omega t} \left[ K + \frac{1}{2} \Big(\delta x(t) + \im \, \delta y(t) \Big) \right] \left(1 + \sqrt{r} \, \ex^{\im (\omega_\IFr t + \theta)} \right) + \xihat_{\inrm}(t) 
\end{equation}
where $r = (P_{lo}/P) \times \omega_{s}/(\omega_s + \omega_{if}) \approx P_{lo}/P \gg 1$ is the ratio between the local oscillator power and the power of the beam used for measurement. The phase $\theta$ is not important here, as the spectrum will not depend on it. Since $\omega_\IFr \gg \omega_\m, \kappa$, the local oscillator does not affect the mechanical oscillator and we can assume that it is promptly reflected. The output mode can be expressed as $\ahat_\mathrm{out}(t) = \ex^{-\im \Omega t} \left(\bar{a}_\mathrm{out}(t) + \dhat_\mathrm{out}(t) \right) $ where $\bar{a}_\mathrm{out}(t)$ describes the average amplitudes of the reflected beams,
\begin{equation}
  \label{eq:OutMean}
  \bar{a}_\mathrm{out}(t) = - K \left( \rho + \sqrt{r} \, \ex^{\im (\omega_\IFr t + \theta)} \right) \ ,
\end{equation}
with $\rho = 1 - \kappa_{\mathrm{ext}}/(\kappa/2 - \im \Delta)$. The first term describes the measurement beam which can be attenuated by the interaction with the cavity if there is internal dissipation, i.e.~if $\kappa_{\mathrm{int}} \neq 0$. The second term describes the promptly reflected local oscillator. The fluctuations around these average amplitudes are given by
\begin{equation}
  \label{eq:OutFluct}
  \dhat_\mathrm{out}(t) = \sqrt{\kappa_\mathrm{ext}} \dhat(t) - \frac{1}{2}\Big(\delta x(t) + \im \, \delta y(t) \Big)  \left(1 + \sqrt{r} \, \ex^{\im (\omega_\IFr t + \theta)} \right) - \xihat_{\inrm}(t) 
\end{equation}
where $\dhat(t)$ is given by Eq.~\eqref{eq:SolutionEOM}. The term proportional to $\sqrt{r}$ is the promptly reflected technical noise in the local oscillator beam.

To calculate the spectrum $S[\omega]$ of the photocurrent $i(t)$, we need to evaluate 
\begin{equation}
  \label{eq:SpectrumDef}
  S[\omega] = \lim_{T \rightarrow \infty} \frac{1}{T} \int_{-T/2}^{T/2} \du t \, \int_{-\infty}^{\infty} \du \tau \, \ex^{\im \omega \tau} \,  \overline{i(t) i(t + \tau)} \ ,
\end{equation}
where the average involves an average over the photoelectron counting distribution \cite{Carmichael1993Open}, which itself is an ensemble average. The current-current correlation function can be expressed by \cite{Carmichael1987Spectrum}
\begin{eqnarray}
  \label{eq:CurrentCorr}
  \overline{i(t) i(t + \tau)} = G^2 \Big( \sigma^2 \langle : \hat{I}(t) \hat{I}(t + \tau) : \rangle + \sigma \langle \hat{I}(t) \rangle \delta(\tau) \Big) \ .
\end{eqnarray}
where $\hat{I}(t) = \ahat^\dagger_\mathrm{out}(t) \ahat_\mathrm{out}(t)$, the colons indicate normal and time ordering, and $\sigma$ is the dimensionless detection efficiency. $G$ is the photodetector gain in units of charge, i.e.~the proportionality constant between the current and the number of photon detections per time. Although $G$ is in general frequency dependent, we will assume that it is approximately constant over an interval of the effective mechanical linewidth $\tilde{\gamma}$. The last term in Eq.~\eqref{eq:CurrentCorr} is due to self-correlation of photoelectric pulses (here we have assumed the detector has infinite bandwidth for simplicity).

The flux operator $\hat{I}(t)$ has many terms, but we are only interested in the beating terms that oscillate at approximately the intermediate frequency $\omega_\IFr$. The noise in $\hat{I}(t)$ at the sidebands $\omega_\IFr \pm \omega_\text{m}$ has two contributions - beating between the average local oscillator beam and the fluctuations in the measurement beam, and beating between the average measurement beam and the noise in the local oscillator beam. Both of these contributions are proportional to $K \sqrt{r}$.

We let $S_\rr[\omega]$ denote the spectrum $S[\omega]$ at the red sideband, i.e.~around the frequency $\omega_\mathrm{r} = \omega_\IFr - \tilde{\omega}_\text{m}$. After a straightforward but tedious derivation, we find 
\begin{equation}
  \label{eq:SpectrumRed}
  S_\rr[\omega] = G_\mathrm{r}^2 \, \sigma \, r K^2 \left[ F_\rr + \frac{\tilde{\gamma}  L_\rr + (\omega - \omega_\mathrm{r}) A_\rr}{(\tilde{\gamma}/2)^2 + (\omega - \omega_\mathrm{r})^2} \right] \ ,
\end{equation}
where we have made the assumption of weak coupling $|\alpha| \ll \kappa$ and $G_\mathrm{r}$ is the gain at frequency $\omega_\mathrm{r}$. The spectrum consists of three terms. The first term is a constant noise floor, whose size is determined by the coefficient
\begin{eqnarray}
  \label{eq:Frr}
  F_\rr & = & 1 + \frac{\sigma}{4} \Big[\left(|\rho|^2 + |\kappa_\mathrm{ext} \chi_\cc[-\omega_\m] - 1|^2 \right) \left(C_{xx} + C_{yy} \right) \notag\\
  & & - 2 \, \mathrm{Re} \big[\rho^\ast \left(\kappa_\mathrm{ext} \chi_\cc[-\omega_\m] - 1\right)\left(C_{xx}  + 2 \im C_{xy} - C_{yy} \right) \big] \Big] \ .
\end{eqnarray}
The first term in \eqref{eq:Frr} is due to shot noise, and the other terms result from technical noise. As a sanity check, we note that for $\kappa_\mathrm{ext} = 0$ or for $|\Delta| \rightarrow \infty$, i.e.~when the measurement beam does not enter the cavity, this coefficient reduces to $F_\rr = 1 + \sigma C_{xx}$. This is independent of phase noise, as it should be since a photodetector cannot detect phase noise directly.

The second term in Eq.~\eqref{eq:SpectrumRed} is a Lorentzian centered on the frequency $\omega_\mathrm{r}$ with a width equal to the mechanical linewidth $\tilde{\gamma}$. The coefficient of this term is
\begin{eqnarray}
  \label{eq:Lrr}
  L_\rr & = & \sigma \kappa_\ext |\alpha|^2 \Big[ |\chi_\cc[-\omega_\m]|^2 (n_\m + 1) + \mathrm{Re} \, \tilde{B}[\omega_\m]\Big]
\end{eqnarray}
with

\begin{eqnarray}
\label{eq:Btilde}
  \tilde{B}[\omega] & = & \frac{\kappa_\ext}{4} |\chi_\cc[-\omega]|^2 \ex^{-\im \phi} \Big[(C_{xx} + \im C_{xy}) B_+[\omega] + (\im C_{xy} - C_{yy}) B_-[\omega]\Big] \notag\\
  & & - \frac{1}{4} \chi^\ast_\cc[-\omega] \ex^{-\im \phi} \Big[ (C_{xx} B_+[\omega] + \im C_{xy} B_-[\omega])(1 + \rho) \notag\\ 
  & & + (\im C_{xy} B_+[\omega] - C_{yy} B_-[\omega])(1 - \rho) \Big] 
\end{eqnarray}
and $B_\pm[\omega] = \ex^{-\im \phi} \chi_\cc[\omega] \pm \ex^{\im \phi} \chi_\cc^\ast[-\omega]$. The first term in (\ref{eq:Lrr}) is the contribution from the mechanical oscillator spectrum, whereas the second originates from optomechanical correlations between the oscillator position and the technical laser noise. 

The third term in the red sideband spectrum Eq.~\eqref{eq:SpectrumRed} is proportional to the imaginary value of the effective mechanical susceptibility and thus changes sign at $\omega_\mathrm{r}$. This antisymmetric term is absent if there is no technical laser noise. Its coefficient is
\begin{eqnarray}
  \label{eq:Arr}
  A_\rr & = & 2 \sigma \kappa_\ext |\alpha|^2 \, \mathrm{Im} \, \tilde{B}[\omega_\m] \ .
\end{eqnarray}

We now move on to the blue sideband at $\omega_\mathrm{b} = \omega_\IFr + \tilde{\omega}_\text{m}$ and denote the spectrum around this frequency by $S_\bb[\omega]$, finding
\begin{equation}
  \label{eq:SpectrumBlue}
  S_\mathrm{bb}[\omega] = G_\mathrm{b}^2 \, \sigma \, r K^2 \left[ F_\bb + \frac{\tilde{\gamma} L_\bb + (\omega - \omega_\mathrm{b}) A_\bb}{(\tilde{\gamma}/2)^2 + (\omega - \omega_\mathrm{b})^2} \right] \ ,
\end{equation}
where $G_\mathrm{b}$ is the photodetector gain at the frequency $\omega_\mathrm{b}$. The spectrum at the blue sideband has the same three terms as the red sideband, but with different coefficients. The noise floor is determined by
\begin{eqnarray}
  \label{eq:Fbb}
  F_\bb & = & 1 + \frac{\sigma}{4} \Big[\left(|\rho|^2 + |\kappa_\ext \chi_\cc[\omega_\m] - 1|^2 \right) \left(C_{xx} + C_{yy} \right) \notag\\
  && - 2 \, \mathrm{Re} \big[\rho^\ast \left(\kappa_\ext \chi_\cc[\omega_\m] - 1\right)\left(C_{xx}  + 2 \im C_{xy} - C_{yy} \right) \big] \Big] \ ,
\end{eqnarray}
the coefficient of the Lorentzian term is
\begin{eqnarray}
  \label{eq:Lbb}
  L_\bb & = & \sigma \kappa_\ext |\alpha|^2 \Big[ |\chi_\cc[\omega_\m]|^2 n_\m - \mathrm{Re} \, \tilde{B}[-\omega_\m]\Big]
\end{eqnarray}
and the coefficient of the antisymmetric term is
\begin{eqnarray}
  \label{eq:Abb}
  A_\bb & = & - 2 \sigma \kappa_\ext |\alpha|^2 \, \mathrm{Im} \, \tilde{B}[-\omega_\m] \ .
\end{eqnarray}

\section{Sideband weights}
\label{sec:SidebandWeights}

Let us define the sideband weights $W_\rr$ and $W_\bb$ as the frequency integral of the spectra $S_\rr[\omega] - S_{0,\rr}$ and $S_\bb[\omega] - S_{0,\bb}$, where $S_{0,\rr}$ and $S_{0,\bb}$ are the noise floors at the red and blue sidebands, respectively. We also assume that the difference in gains at the red and blue sidebands are accounted for. The antisymmetric parts proportional to $A_\rr$ and $A_\bb$ will not contribute to the integral, and we find that the ratio of the sideband weights is
\begin{equation}
  \label{eq:RatioSidebandWeights}
  \frac{W_\bb}{W_\rr} =  \frac{ |\chi_\cc[\omega_\m]|^2 n_\m - \mathrm{Re} \, \tilde{B}[-\omega_\m]  }{  |\chi_\cc[-\omega_\m]|^2 (n_\m + 1) + \mathrm{Re} \, \tilde{B}[\omega_\m] } \ .
\end{equation}
In the absence of technical laser noise, and at zero detuning $\Delta = 0$, this reduces to the Boltzmann weight, $W_\bb/W_\rr =  n_\m/(n_\m + 1)$, as is well known \cite{Rae2007Theory,Marquardt2007Quantum}. In general, however, the ratio of the sideband weights do not provide a direct measure of the effective phonon number $n_\m$. To determine $n_\m$ by this method, one needs to know the detuning $\Delta$, the decay rates $\kappa, \kappa_\ext$, and the noise coefficients $C_{xx}$ etc.~to a sufficient accuracy. 

To illustrate that one needs to be careful in this regard, let us for a moment assume that $\kappa_\mathrm{int} = \Delta = 0$ and that phase noise dominates, i.e.~$C_{xx} \ll C_{xy} , C_{yy}$. This gives
\begin{equation}
  \label{eq:RatioExample}
  \frac{W_\bb}{W_\rr} = \frac{n_\m + C_{xy} |\chi_\cc[\omega_\m]|^2 \kappa \omega_\m/2 }{n_\m + 1 + C_{xy} |\chi_\cc[\omega_\m]|^2 \kappa \omega_\m/2} = \frac{n_\mathrm{est}}{n_\mathrm{est} + 1} \ ,
\end{equation}
such that one would naively estimate the average phonon number to be $n_\mathrm{est} = n_\m + C_{xy} |\chi_\cc[\omega_\m]|^2 \kappa \omega_\m/2$ if technical noise is neglected. We see that, since the cross-correlation coefficient $C_{xy}$ can be negative, this can potentially lead to underestimating the phonon number. Note also that the absence of the phase noise coefficient $C_{yy}$ in this simple example crucially depends on the assumption of exactly zero detuning.

\begin{figure}
  \includegraphics[width=\textwidth]{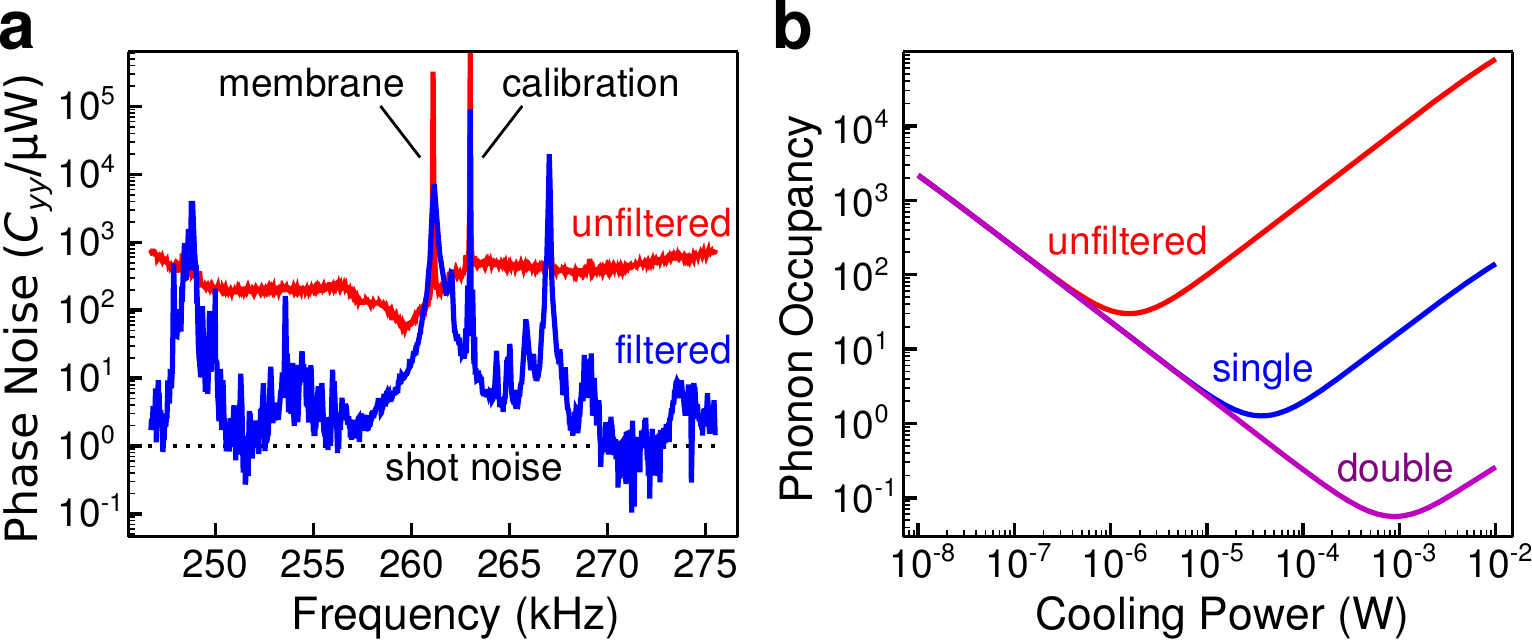}
  \caption{Laser noise and cooling limits. \textbf{a} Classical phase noise $C_{yy}$ of our cooling laser for $P_\text{p}$ = 1 \textmu W measured using the cold cavity as a reference, both with (blue) and without (red) the filter cavity. Near $\omega_\text{m}$, the unfiltered noise background corresponds to $C_{yy} = 200$~at 1 \textmu W. At nearby frequencies (e.g. 270 kHz) the filter cavity performs as expected, but membrane vibrations (at 261 kHz) and other technical noise added by our system clouds the measurement of $C_{yy}$ at other frequencies. The large peak at 263 kHz corresponds to an intentional, known phase modulation (applied with an EOM) that we use as a reference to calibrate this data. The unfiltered data was taken with $P_\text{lo} = 423$~\textmu W and $P_\text{p} = $ 1.5 \textmu W. The filtered data was taken with $P_\text{lo} = $~239 \textmu W, $P_\text{p} =$ 16.3 \textmu W. \textbf{b} Predicted phonon occupancy versus cooling laser power for zero (red), one (blue), and two (purple) passes through the filter cavity described in the text.}
  \label{fig3}
\end{figure}

\section{Discussion}

The above analysis makes it clear that in order to reliably perform a calibrated heterodyne thermometry measurement, we must first develop a reliable characterization of the laser's classical noise. We have made some initial estimates using the experimental apparatus described above. 

It is straightforward to determine the amplitude noise $C_{xx}$ by directly measuring laser power fluctuations with a photodiode (and subtracting the shot noise and the photodiode's dark noise) \cite{Yang2011Progress}. For our cooling laser, this yields a value $C_{xx}$ = 0.02 for laser power $P_p = 1$ \textmu W. 

We can estimate the phase noise $C_{yy}$ by using the optical circuit described above, and allowing the membrane cavity to serve as a reference. We do this by comparing the noise spectra of the laser light leaving the cryostat under two conditions: with the laser tuned far from the cavity resonance (so the signal photodiode is only sensitive to amplitude noise) and with the laser near resonance (so phase noise is converted to amplitude noise) \cite{Yang2011Progress}. Figure \ref{fig3}a shows a plot of the cooling laser's phase noise near $\omega_\text{m}$. The ``unfiltered'' (red) spectrum corresponds to the free-running cooling laser used in the experiment. A peak from the membrane's thermal motion, along with a known phase modulation peak at 263 kHz (use to calibrate this data), sits on top of a broad background arising from the cooling laser's intrinsic phase noise of $C_{yy} \approx 200$ at 1 \textmu W near $\omega_\text{m}$. The estimate shown in Fig. \ref{fig3}a assumes $C_{xy} = 0$ for simplicity, though letting $C_{xy}$ vary over the allowed range $\pm \sqrt{C_{xx}C_{yy}}$ only changes this estimate by a few percent. 

Given this estimate of the cooling laser's classical noise, we can estimate the fundamental limits of laser cooling with this system using Eq. \ref{eq:19} above. The curve labeled ``unfiltered'' in Fig. \ref{fig3}b shows the expected average phonon occupancy as a function of power for the free-running cooling laser. Also included in this calculation is a 1.5 \textmu W signal laser with $\Delta_\text{s} = 0$, $C_{xx} = 0.13$, $C_{xy} = 0$, and $C_{yy} = 780$. These values of $C_{xx}$ and $C_{yy}$ correspond to similar measurements of the signal laser, and we again assume $C_{xy} \approx 0$ (the result in Fig. \ref{fig3}b is insensitive to the value of $C_{xy}$). The minimum phonon occupancy that could be achieved with the current cryogenic apparatus is $\sim 30$, corresponding to a temperature $\sim$ 375 \textmu K.

In an effort to reduce the classical noise, we have inserted a filter cavity in the cooling laser's room-temperature beam path. This cavity has a resonance width $\kappa_\text{filter}/2\pi$ = 22 kHz, meaning the cooling laser's classical noise power should scale down by a factor $1 + 4\omega_\text{m}^2/\kappa_\text{filter}^2 \sim 500$. We lock the filter cavity to the free-running cooling laser and measure its noise again as shown in Fig. \ref{fig3}a. We observe the expected reduction at some frequencies near $\omega_\text{m}$ (e.g. 270 kHz), and attribute the remaining noise structure to our use of the acoustically-sensitive membrane cavity as the measurement reference. Nonetheless, the observation of filtered laser noise while locked to the cryogenic cavity is encouraging, and we expect the filter cavity to perform as predicted over the full spectrum in a vibration-isolated system. 

Once the filter cavity is locked to the cooling laser, it is straightforward to rotate the polarization of the output light and pass it through the filter cavity again with no additional feedback \cite{Hald2005Efficient}. This enables four poles of passive filtering, and would further reduce the cooling laser noise. Such a double-filtered cooling laser would allow the membrane to be laser cooled very close to its quantum mechanical ground state, as shown in Fig. \ref{fig3}b.

\section{Acknowledgments}

The authors acknowledge support from AFOSR (No. FA9550-90-1-0484), NSF 0855455, NSF 0653377, and NSF DMR-1004406. KB acknowledges financial support from The Research Council of Norway and from the Danish Council for Independent Research under the Sapere Aude program. The authors would also like to acknowledge helpful conversations and technical support from N. Flowers-Jacobs.

\section{References}

\bibliography{JackSankey}{}
\bibliographystyle{unsrt} 
\end{document}